\documentclass[12pt]{article}
\usepackage{amsmath,amssymb,bm,graphicx,epsfig}
\usepackage[sort&compress,numbers]{natbib}
\usepackage{lineno,booktabs,url}
\usepackage{subfigure}
\RequirePackage[colorlinks,citecolor=blue,urlcolor=blue,linkcolor=blue]{hyperref}

\newcommand\as{\alpha_S} 
\def\ltap{\raisebox{-.6ex}{\rlap{$\,\sim\,$}} \raisebox{.4ex}{$\,<\,$}} 
\def\gtap{\raisebox{-.6ex}{\rlap{$\,\sim\,$}} \raisebox{.4ex}{$\,>\,$}} 

\def\nn{\nonumber}

\def\beqn{\begin{eqnarray}} 
\def\eeqn{\end{eqnarray}} 
\def\beq{\begin{equation}} 
\def\eeq{\end{equation}}

\newcommand\f[2]{\frac{#1}{#2}} 
\def\la{\lambda}

\begin{document}
\include{epsf} 
\begin{titlepage}
\begin{flushright}
IFIC/23-11\\
FTUV-23-0321.0606
\end{flushright}

\renewcommand{\thefootnote}{\fnsymbol{footnote}}
\vspace*{2.cm}

\begin{center}
{\Large \bf 
  Drell--Yan lepton-pair production: 
  \\[0.1cm]
  $\bf q_T$ resummation at approximate N$\bf ^4$LL+N$\bf ^4$LO  accuracy \\[0.1cm]
}
\end{center}

\par \vspace{2mm}
\begin{center}
  {\bf Stefano Camarda${}^{(a)}$},  {\bf Leandro Cieri${}^{(b)}$}
  and {\bf Giancarlo Ferrera${}^{(c)}$}\\

\vspace{5mm}

${}^{(a)}$ 
CERN, CH-1211 Geneva, Switzerland\\\vspace{1mm}

${}^{(b)}$ 
Instituto de F\'isica Corpuscular, Universitat de Val\`encia - Consejo Superior de
Investigaciones Cient\'ificas, Parc Cient\'ific, E-46980 Paterna, Valencia, Spain\\\vspace{1mm}

${}^{(c)}$ 
Dipartimento di Fisica, Universit\`a di Milano and\\ INFN, Sezione di Milano,
I-20133 Milan, Italy\\\vspace{1mm}

\end{center}

\vspace{1.5cm}

\par \vspace{2mm}
\begin{center} {\large \bf Abstract} \end{center}
\begin{quote}
\pretolerance 10000
We consider  Drell--Yan lepton pairs produced
in hadronic collisions.
We present   high-accuracy QCD predictions
for the transverse-momentum ($q_T$) distribution
and fiducial cross sections in the small $q_T$ region.
We resum to all perturbative orders the
logarithmically enhanced contributions
up to the
next-to-next-to-next-to-next-to-leading logarithmic (N$^4$LL) accuracy
and we include the  hard-virtual coefficient at the
next-to-next-to-next-to-leading order (N$^3$LO)
(i.e.\ $\mathcal{O}(\alpha_S^3)$) with an approximation of the N$^4$LO
coefficients.
The massive axial-vector and vector contributions up to three loops have also been
consistently included. 
The resummed partonic cross section
is convoluted with approximate N$^3$LO parton distribution functions.
We show numerical results at LHC energies of
resummed $q_T$ distributions for
$Z/\gamma^*, W^\pm$  production and decay, including the $W^\pm$ and $Z/\gamma^*$  ratio, estimating the corresponding
uncertainties from missing higher orders corrections and
from incomplete or missing perturbative information coefficients
at N$^4$LL and N$^4$LO.
Our resummed calculation has been encoded in the public
numerical program {\ttfamily DYTurbo}.

\end{quote}

\vspace*{\fill}
\vspace*{2.5cm}

\begin{flushleft}
March 2023
\end{flushleft}
\end{titlepage}

\setcounter{footnote}{1}
\renewcommand{\thefootnote}{\fnsymbol{footnote}}

The production of high invariant mass ($M$) lepton pairs in hadronic collision,
through the Drell--Yan (DY) mechanism\,\cite{Drell:1970wh,Christenson:1970um},
is extremely important for physics studies at hadron colliders and attracted
a great deal of attention from the experimental and theory communities.
Since the early days of QCD remarkable efforts have been devoted to detailed calculations of
the dominant QCD higher-order radiative corrections 
of fiducial cross sections and kinematical distributions.

A sufficiently inclusive cross section can be  perturbatively computable as an expansion
in the QCD coupling $\alpha_S=\alpha_S(\mu_R^2)$ where the normalization scale $\mu_R$ is of the order of the invariant mass
$M$.
However the bulk of experimental data lies in the small transverse momentum ($q_T$) region $q_T\ll M$ where the fixed-order expansion 
is spoiled by the presence of enhanced logarithmic corrections, $\alpha_S^n  \ln^m(M^2 /q_T^2 )$ 
 of soft and collinear origin.
 In order to obtain reliable predictions, these logarithmic terms have to be systematically resummed to all orders in perturbation
 theory\,\cite{Dokshitzer:1978yd,Parisi:1979se,Collins:1984kg} (resummed calculation and studies applying different formalism and various levels of theoretical accuracy
have been performed in Refs.\,\cite{Bozzi:2005wk,Bozzi:2008bb,Bozzi:2010xn,Catani:2010pd,Becher:2010tm,Becher:2011xn,Collins:2011zzd,Collins:2012uy,Banfi:2012du,Guzzi:2013aja,Collins:2014jpa,Catani:2015vma,Ebert:2016gcn,Coradeschi:2017zzw,Scimemi:2017etj,Bizon:2018foh,Bizon:2019zgf,Becher:2019bnm,Bertone:2019nxa,Bacchetta:2019sam,Ebert:2020dfc,Becher:2020ugp,Re:2021con,Alioli:2021qbf,Ju:2021lah,Neumann:2022lft,Chen:2022cgv}.

In this paper we consider the Drell--Yan lepton pair production in the small $q_T$ region and
we apply the QCD transverse-momentum resummation formalism developed in
Refs.\,\cite{Bozzi:2005wk,Bozzi:2010xn,Catani:2015vma}.
We resum all the
logarithmically enhanced contributions
up to the next-to-next-to-next-to-next-to-leading logarithmic (N$^4$LL) accuracy
and we include the  hard-virtual coefficient at the
next-to-next-to-next-to-leading order (N$^3$LO)
(i.e.\ $\mathcal{O}(\alpha_S^3)$) with an estimate of the N$^4$LO effects.

In the  $Z$ boson case, because of the axial coupling, Feynman diagrams with quark loops
contribute to the cross-section at $\mathcal{O}(\alpha_S^2)$ and  $\mathcal{O}(\alpha_S^3)$.
These contributions, also known as singlet contributions, cancel out for each isospin multiplet when massless quarks are considered.
The effect of a finite top-quark mass in the third generation has been considered
at $\mathcal{O}(\alpha_S^2)$
in Refs.\,\cite{Dicus:1985wx,Rijken:1995gi} and has been found extremely small compared to the NNLO corrections. 
However these effects are not completely negligible when compared to the N$^3$LO corrections\,\cite{Ju:2021lah}. We have considered
the effect of a finite top-quark mass including in our calculation the singlet contributions up to
$\mathcal{O}(\alpha_S^3)$ by using the calculation of the quark axial form factor in QCD up to three loops\,\cite{Chen:2021rft}.
We consistently  included also the quark-loop mediated three-loop singlet corrections which contribute, via vector coupling, both to $Z$ and $\gamma^*$ production at $\mathcal{O}(\alpha_S^3)$ \,\cite{Lee:2010cga,Gehrmann:2010ue}

At large value of  $q_T$ ($q_T\sim M$) fixed-order perturbative expansion is fully justified. In this region, the QCD radiative corrections
are known up to $\mathcal{O}(\alpha_S^3)$  numerically through the fully exclusive NNLO calculation of vector boson production in association with jets\,\cite{Boughezal:2015dva,Gehrmann-DeRidder:2015wbt,Boughezal:2015ded,Boughezal:2016dtm,Boughezal:2016isb,Gehrmann-DeRidder:2016cdi,Gehrmann-DeRidder:2016jns,Gehrmann-DeRidder:2017mvr,Neumann:2022lft}.
In particular the calculation of $Z+jet$ production at NNLO has been encoded in the public code MCFM\,\cite{Neumann:2022lft}.  
Resummed and fixed-order calculation have to be consistently (i.e.\ avoiding double counting) matched 
at intermediate values of $q_T$ in order to obtain theoretical predictions with uniform accuracy over the entire range of $q_T$.

Our resummed calculation for $Z/\gamma^*$ and $W^\pm$  production and decay up to approximated N$^4$LL+N$^4$LO accuracy,
together with the asymptotic expansion up to $\mathcal{O}(\alpha_S^3)$,
has been  implemented in the public
numerical program {\ttfamily DYTurbo}\,\cite{Camarda:2019zyx,dyturbo}  which provides fast and
numerically precise predictions
including the full kinematical dependence of the decaying lepton pair
with the corresponding spin correlations and the finite
value of the $Z$ boson width.

In this paper we are focusing on the impact of the N$^4$LL resummed logarithmic terms. We thus consider only the the small $q_T$ region and we
not include the matching with fixed-order predictions
which can be implemented starting from the results of Refs.\,\cite{Boughezal:2015dva,Gehrmann-DeRidder:2015wbt,Boughezal:2015ded,Boughezal:2016dtm,Boughezal:2016isb,Gehrmann-DeRidder:2016cdi,Gehrmann-DeRidder:2016jns,Gehrmann-DeRidder:2017mvr,Neumann:2022lft}
and subtracting the {\itshape asymptotic} expansion of the resummed calculation at the same perturbative order as encoded in {\ttfamily DYTurbo}.
Resummed results at N$^3$LL +N$^3$LO  matched with the NNLO calculation at large $q_T$ have been presented
in Refs.\,\cite{Camarda:2021ict}. 
Here we extend the results of Ref.\,\cite{Camarda:2021ict} by extending the resummation accuracy at approximated N$^4$LL+N$^4$LO and
by presenting results for $W^\pm$ boson production and decay.
A brief review of the resummation formalism of Refs.\,\cite{Bozzi:2005wk,Bozzi:2010xn,Catani:2015vma}
is given in Appendix\,\ref{appendix} together  
with a collection of the numerical coefficients needed at N$^4$LL+N$^4$LO accuracy.

In the following we consider  $Z/\gamma^*,W^\pm$ production and leptonic decay at the Large Hadron Collider (LHC). 
We present resummed predictions up to N$^4$LL accuracy
including the hard-virtual coefficient up to N$^3$LO together with an approximation of the N$^4$LO ones.
The hadronic cross section is obtained by convoluting
the partonic cross section in Eq.\,(\ref{partXS2}) with the parton densities functions
(PDFs) from MSHT20aN3LO 
 set\,\cite{McGowan:2022nag} at
the approximate N$^3$LO  with $\as(m_Z^2)=0.118$
where we have evaluated $\as(\mu_R^2)$ at $(n\!+\!1)$-loop order
at N$^n$LL accuracy.
We use the so called $G_\mu$ scheme for EW couplings
with input parameters 
$G_F = 1.1663787\times 10^{-5}$~GeV$^{-2}$,
$m_Z = 91.1876$~GeV, $\Gamma_Z=2.4952$~GeV, $m_W = 80.379$~GeV, $\Gamma_W=2.091$~GeV.
In the case of $W$ production, we use the following CKM matrix elements: $V_{ud} = 0.97427$, $V_{us} = 0.2253$,
$V_{ub} = 0.00351$, $V_{cd} = 0.2252$, $V_{cs} = 0.97344$, $V_{cb} = 0.0412$.
We work with $N_f=5$ massless quarks and we use $m_{top}=173$~GeV for the top-loop mediated singlet contributions. 
Our calculation implements 
the leptonic decays $Z/\gamma^* \to l^+l^-$, $W\pm \to l \nu$ 
and we include the effects of the $Z/\gamma^*$ interference and of the
finite widths 
of the $W$ and
$Z$ boson
with the corresponding spin correlations and the full dependence  
on the kinematical variables of final state leptons.
This allows us to take into account the typical 
kinematical cuts on final state leptons that are
considered in the experimental analysis. 
The resummed calculation  at fixed lepton momenta requires a $q_T$-recoil procedure.
We implement the general procedure described in Ref.\,\cite{Catani:2015vma}
which is equivalent to compute the Born level distribution 
$d{\sigma}^{(0)}$ of Eq.\,(\ref{resum})
in the Collins--Soper rest frame\,\cite{Collins:1977iv}.

As for the non-perturbative (NP) 
effects at very small transverse momenta we introduced, in the conjugated $b$-space, 
a NP form factor of the form\,\cite{Collins:2014jpa}
\begin{equation}
S_{NP}(b)=\exp\{-g_1 b^2-g_{K}(b)\,\ln(M^2/Q_0^2)\}  
\end{equation}
where
\begin{equation} 
g_{K}(b)=g_0 \left(1-\exp\left[-\frac{C_F \alpha_S((b_0/b_\star)^2)b^2}{\pi g_0 b_{\textrm{lim}}^2}\right]\right)\,,
\end{equation} 
with $g_1=0.5$~GeV$^2$, $Q_0=1$~GeV, $g_0=0.3$, $b_{\textrm{lim}}=1.5$~GeV$^{-1}$ and
\begin{equation}
\label{bstar}
b_\star^2=b^2b_{\textrm{lim}}^2/(b^2+b_{\textrm{lim}}^2)\,.
\end{equation} 
The variable $b_\star$
is also used to regularize the perturbative form factor at very large value of $b$ ($b\gtap 1 /\Lambda_{QCD}$, where $\Lambda_{QCD}$ is the 
scale of the Landau pole of the perturbative running coupling $\alpha_S(q^2)$) which correspond to very small values of $q_T$
($q_T\ltap \Lambda_{QCD}$) through the so-called `$b_\star$ 
prescription'~\cite{Collins:1981va,Collins:1984kg} which consist in
the {\itshape freezing} of the integration over $b$ 
below the upper limit $b_{\textrm{lim}}$ through
the replacement $b\to b_\star$. 
An alternative regularization procedure of the Landau singularity,
which have also been implemented in the
{\tt DYTurbo} numerical program,
is the so-called
Minimal Prescription \cite{Catani:1996yz,Laenen:2000de,Kulesza:2002rh}.

We have thus considered the production of
$l^+ l^-$ pairs from  $Z/\gamma^*$ decay 
at the LHC ($\sqrt{s}=13\,$TeV) with the following fiducial cuts:
the leptons are required to have transverse momentum $p_T>25\,$GeV,
pseudo-rapidity $|\eta|<2.5$ while the lepton pair system is required
to have  
an invariant mass of $80<M_{l^+ l^-}<100\,$GeV
with transverse momentum $q_T<30$~GeV.

In order
to estimate the size of yet uncalculated higher-order terms
and the ensuing perturbative uncertainties we consider
the  dependence of the results from the
auxiliary scales $\mu_F$, $\mu_R$ and $Q$.
We thus perform an independent variation of 
$\mu_F$, $\mu_R$ and $Q$
in the range
$M/2 \leq \{ \mu_F, \mu_R, Q \} \leq 2M$ with the constraints
$0.5\leq \{ \mu_F/\mu_R, Q/\mu_R,
Q/\mu_F \}\leq 2$.

\begin{figure}[t]
\begin{center}
  \includegraphics[width=0.495\textwidth]{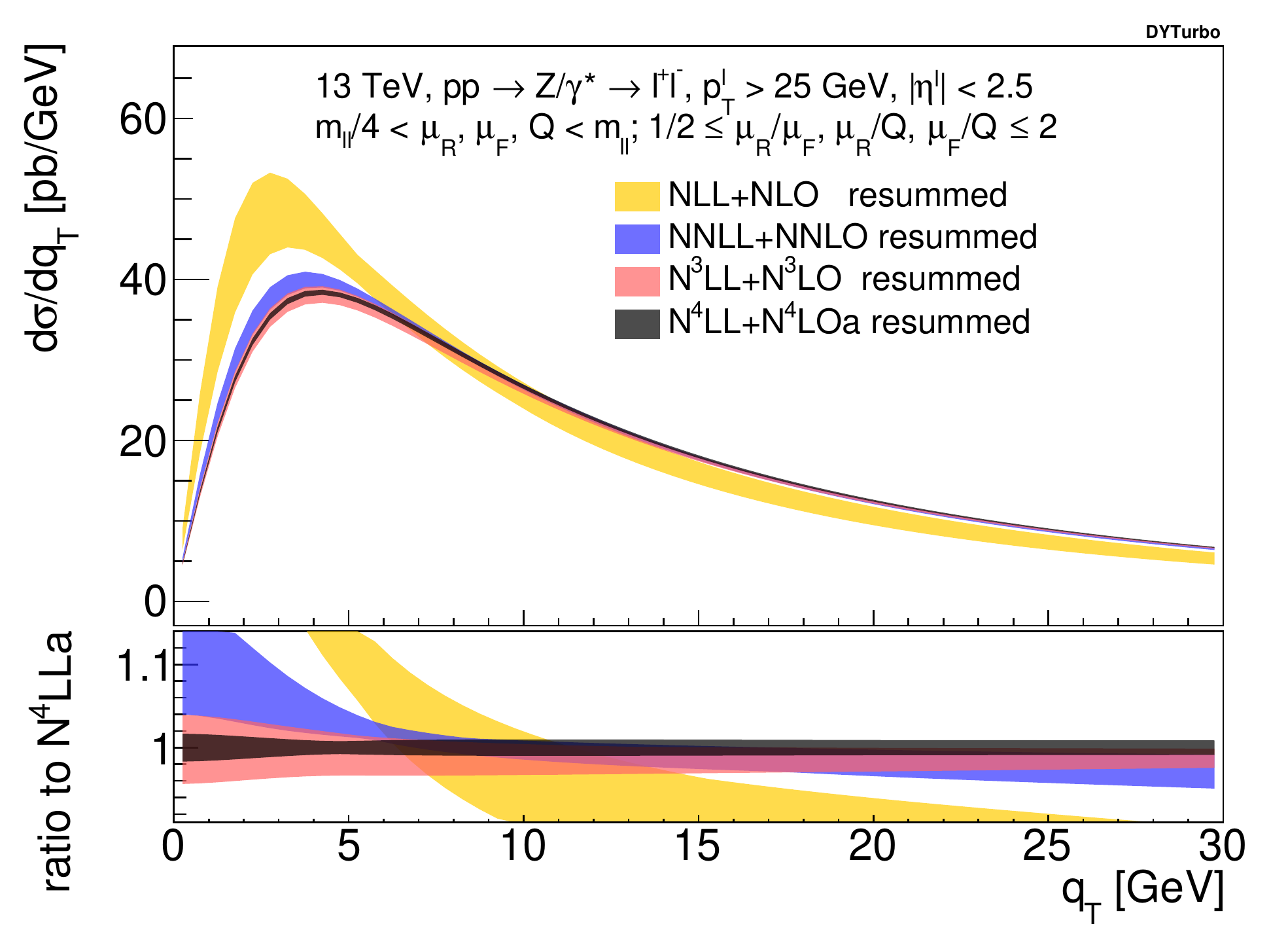}
  \includegraphics[width=0.495\textwidth]{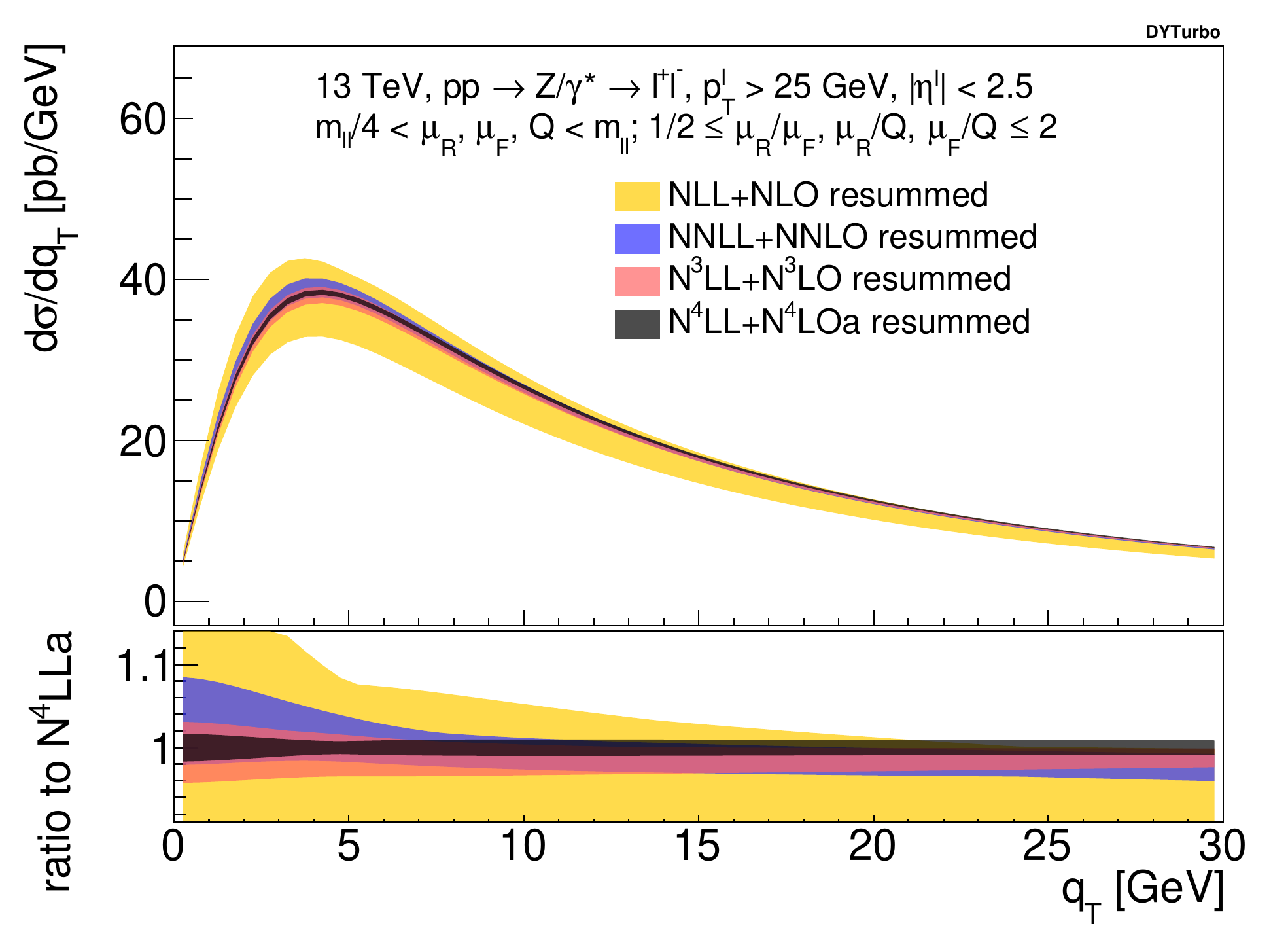}
  
\end{center}
\caption{
  \label{fig1}
  {\em
    The $q_T$ spectrum of $Z/\gamma^*$ bosons with lepton selection cuts at  the LHC ($\sqrt{s}=13$~TeV)
    at various perturbative orders.
    Resummed component (see Eq.\,(\ref{partXS2})) of the hadronic cross-section
with scale variation bands as  defined in the text. The order of the parton density evolution is set consistently with the order of the resummation (left) or with the order of the PDFs (right).
}}
\end{figure}

In Fig.\,\ref{fig1} we consider  $Z/\gamma^*$ production and decay and we show the
resummed component (see Eq.\,(\ref{partXS2})) of the transverse-momentum
distribution in the small-$q_T$ region.
The label N$^n$LL+N$^n$LO ($n=1,2,3$) indicates that we perform the resummation
of logarithmic enhanced contribution at  N$^n$LL accuracy
including the hard-virtual coefficient at  N$^n$LO while the label N$^4$LL+N$^4$LOa indicate that
we perform the resummation at N$^4$LL accuracy with
the hard-virtual coefficient at N$^4$LO and an estimate of yet
not known N$^4$LO
corrections\,\footnote{Incidentally we observe that our prediction at N$^4$LL+N$^4$LOa include the full perturbative information
  contained in the so called N$^4$LL  accuracy and also a reliable approximation of the
N$^4$LL' accuracy as sometimes defined in the literature.}.

In the left panel of Fig.\,\ref{fig1} we show the resummed predictions
following the original formalism of Refs.\,\cite{Bozzi:2005wk,Bozzi:2010xn,Catani:2015vma}.
The lower panel shows the ratio of the distribution with respect to the N$^4$LLa prediction at the central value of the scales $\mu_F=\mu_R=Q= M$.
We observe that the NLL+NLO and NNLL+NNLO scale dependence bands 
do not overlap thus showing that the NLL+NLO scale variation underestimates the true perturbative uncertainty. 
This feature was already observed and discussed in Refs.\cite{Catani:2015vma,Camarda:2021ict}.
In the present case the lack of overlap can be ascribed to the fact
that we are using  the same N$^3$LO parton densities set at NLL, NNLL, N$^3$LL and N$^4$LL accuracy. This choice introduce a formal mismatch
between the N$^3$LO Altarelli-Parisi evolution as  encoded in the
N$^3$LO parton densities functions and the corresponding
N$^k$LO evolution included in the N$^{k+1}$LL partonic resummed formula.

In order to show that this is indeed the case, in the right panel of Fig.\,\ref{fig1} we show the resummed predictions in which we set the order of Altarelli-Parisi evolution in the resummed prediction to be equal to the 
order of the parton densities (i.e.\ both at approximated N$^3$LO). 
In practice, with this choice, we are modifying the  NLL, NNLL and N$^3$LL predictions by including formally subleading logarithmic 
corrections\,\footnote{We note that this inclusion of formally subleading terms is similar to what happen in the the Collins, Soper and Sterman resummation formalism\,\cite{Collins:1984kg} where the parton densities are evaluated at the scale $b_0/b$\,\cite{Parisi:1979se}.}. 
We observe that with this choice the scale dependence bands show a nice overlap at subsequent orders thus indicating that the lack of overlap of the previous case is indeed related to the mismatch in the order of the evolution of parton densities. 
However we also note that by keeping fixed the evolution of the parton densities at subsequent orders inevitably underestimates the impact of higher order corrections included in the PDFs.

Finally, we observe that the choice of the order in the the evolution of parton densities only affects the NLL+NLO and, with a minor extent, NNLO+NNLO theoretical predictions and corresponding uncertainties. Its impact is negligible at N$^3$LL+N$^3$LO (the N$^4$LL+N$^4$LOa prediction is independent by the choice). 
Since we are mainly interested on the impact of N$^4$LL+N$^4$LOa corrections with respect to the N$^3$LL+N$^3$LO results in the following we show numerical results only for the case in which the order of evolution of parton densities is set consistently with the order of the PDF set.

In both the left and right panel of Fig.\,\ref{fig1} the scale dependence is consistently reduced increasing the perturbative order, in particular it is roughly reduced by a factor of 2 going from N$^3$LL to N$^4$LLa. 
The scale variation at  N$^4$LLa accuracy is around $\pm 1.5\%$  at  $q_T\sim 1\,$GeV, then it reduces at $\pm 1\%$  level at the peak ($q_T\sim 4\,$GeV) and  remains roughly constant up to $q_T\sim 30\,$GeV. 

In the results of Fig.\,\ref{fig1} we considered the 
effect of a finite top-quark mass including the singlet contributions mediated
by heavy-quark loops at NNLO and N$^3$LO. As already found in the literature \cite{Dicus:1985wx,Rijken:1995gi} the impact of these contribution
is extremely small, the effect is of $-0.04\%$ at NNLO and less than $+0.001\%$ at N$^3$LO.

\begin{figure}[t]
\begin{center}
  \includegraphics[width=0.7\textwidth]{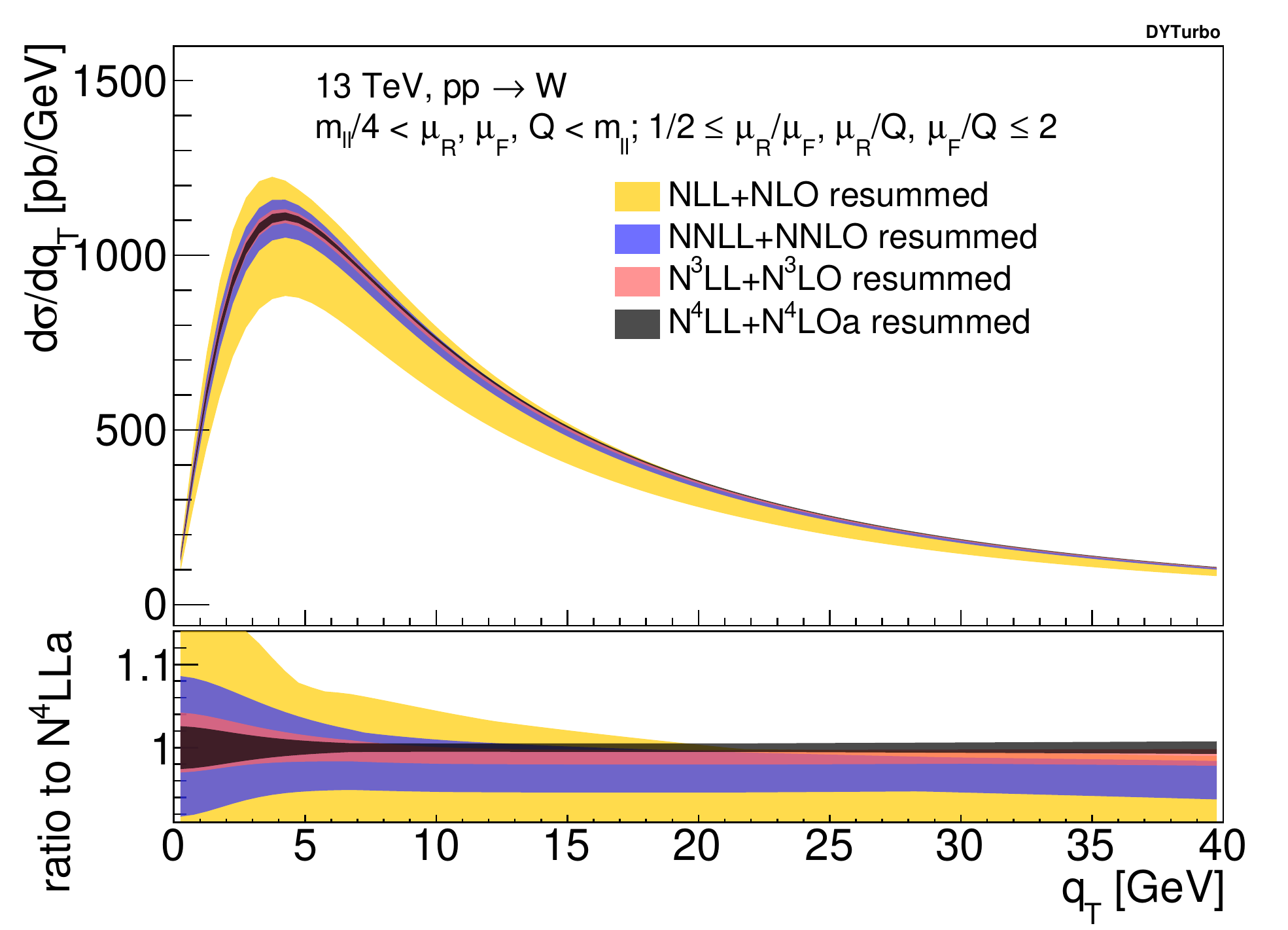}
  
\end{center}
\caption{
  \label{fig2}
  {\em
    The $q_T$ spectrum of $W^+$ and $W^-$ bosons 
    with inclusive leptonic decay at the LHC ($\sqrt{s}=13$~TeV)
    at various perturbative orders.
    Resummed component (see Eq.\,(\ref{partXS2})) of the hadronic cross-section
with scale variation bands as  defined in the text. 
}}
\end{figure}

In Fig.\,\ref{fig2} we consider $W$ boson production and decay into a $l \nu_l$  pair showing the
resummed component of the transverse-momentum
distribution in the small-$q_T$ region at different perturbative orders. In this
case we do not consider kinematical selection cuts apart a lower limit of $50$\,GeV  on the invariant mass of the vector boson (lepton pair) which
is necessary in order to fix a hard scale for the process.
Also in this case we observe that the scale dependence is consistently reduced increasing the perturbative order.
The scale variation at  N$^4$LLa accuracy is around $\pm 2\%$  at  $q_T\sim 1\,$GeV,
then it reduces at $\pm 1\%$  level at the peak ($q_T\sim 4\,$GeV), it further decrease to
$\pm 0.5\%$ for $q_T\sim 7\,$GeV and remains below $\pm 1\%$ level up to $q_T\sim 30\,$GeV. 

The knowledge of the shape of the $W$ boson $q_T$ distribution 
and its uncertainty is particularly
important since it affects the measurement of the $W$ mass. 
However the $W$ boson $q_T$ spectrum  is not directly experimental accessible with
good resolution due to the neutrino in final state of the leptonic $W$ decay.  
Conversely, the $q_T$ spectrum of the $Z$ boson has been measured with great precision. 
Therefore a precise theoretical prediction of the ratio of $W$ and $Z$ $q_T$ distributions, together with the measurement of the 
$Z$ boson $q_T$ spectrum, gives stringent information on the $W$ spectrum. 

\begin{figure}[t]
 \begin{center}
    \subfigure[]{\includegraphics[width=0.495\textwidth]{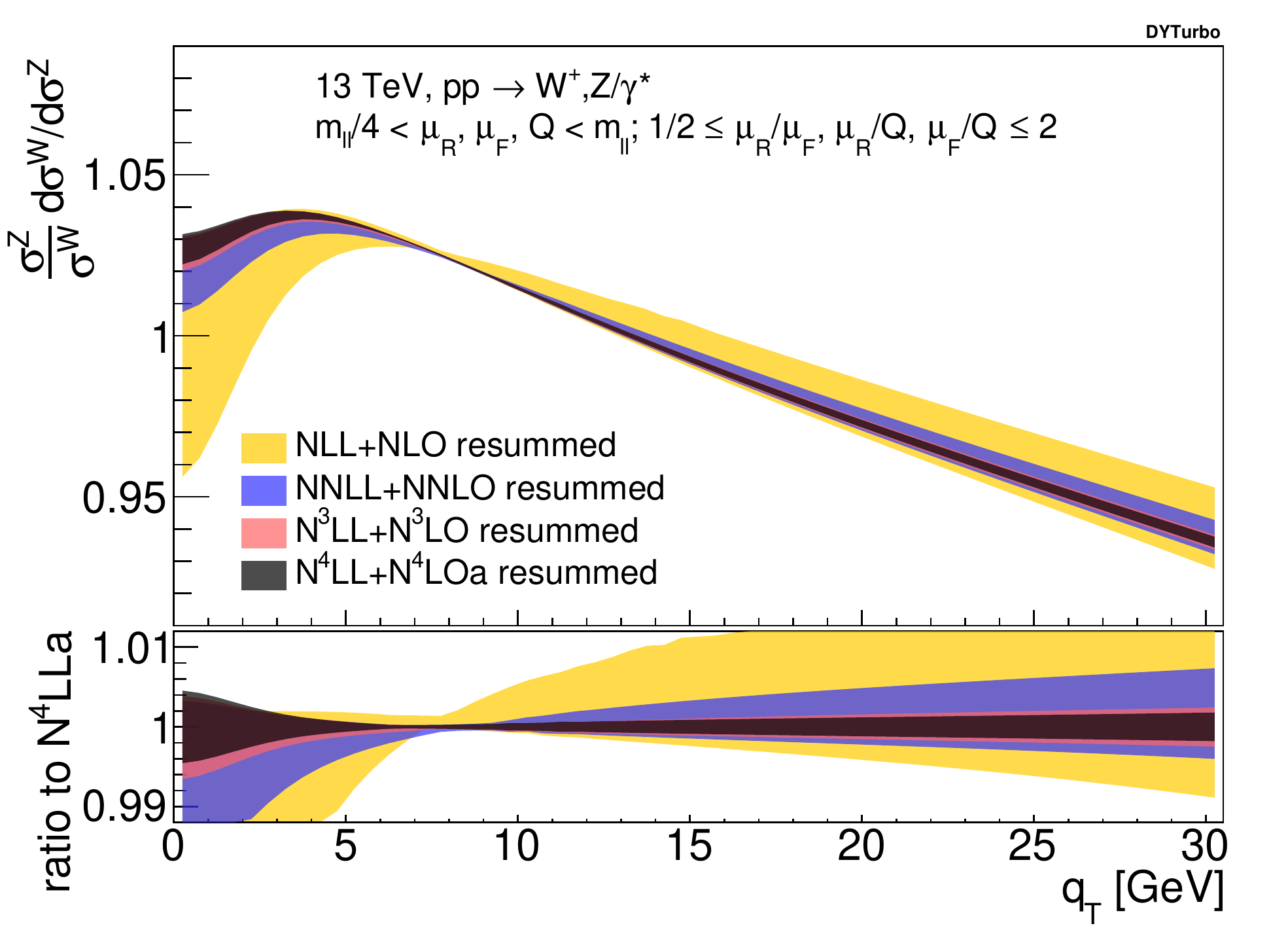}} 
    \subfigure[]{\includegraphics[width=0.495\textwidth]{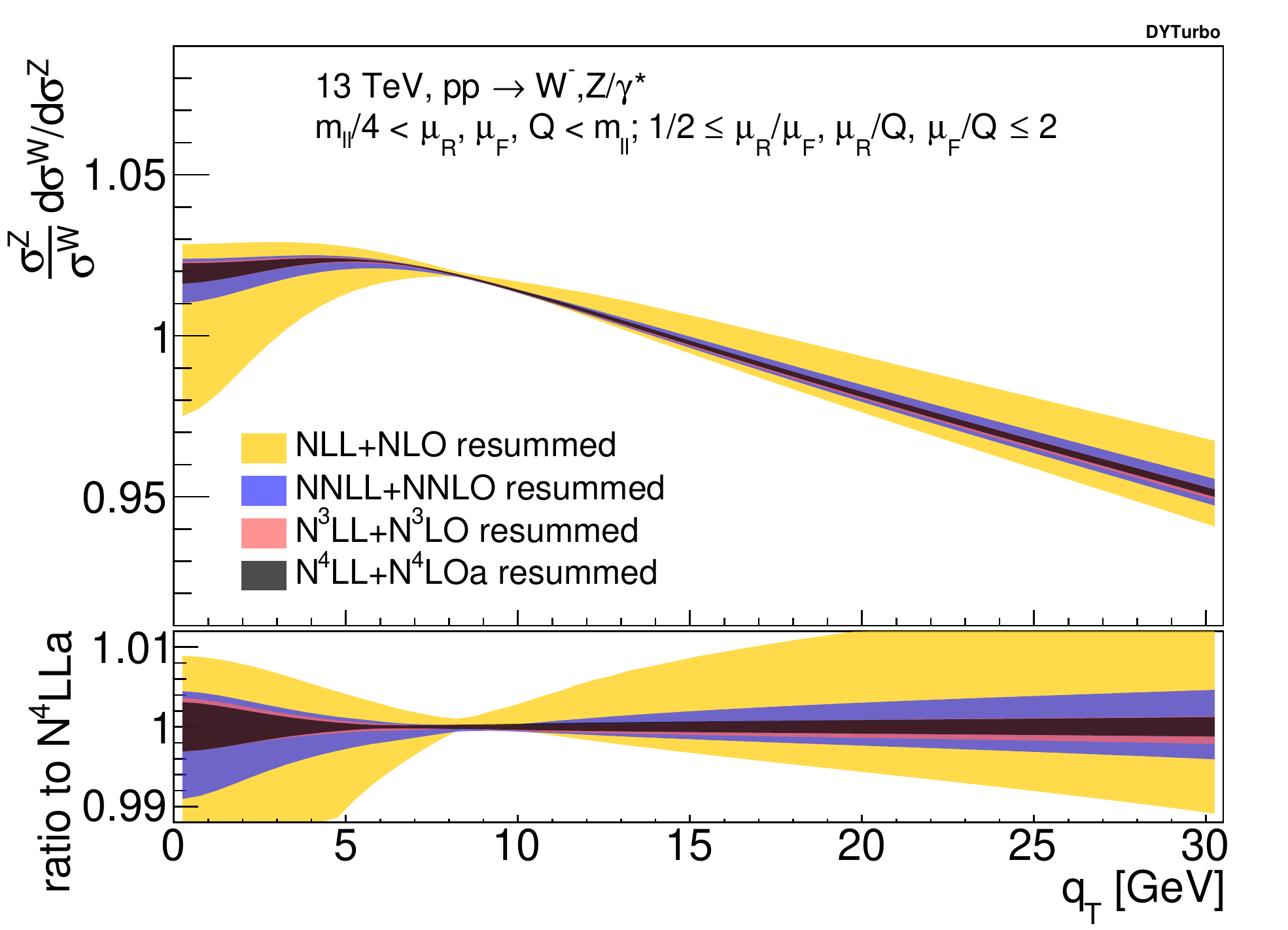}}
 \end{center}
\caption{
  \label{fig3}
  {\em
    The normalized ratio of $q_T$ spectra of $W$ and $Z/\gamma^*$ bosons at the LHC ($\sqrt{s}=13$~TeV)
    at various perturbative orders for $W^+/Z$ (left) and $W^-/Z$ (right).
    Resummed component (see Eq.\,(\ref{partXS2})) of the hadronic cross-section
with scale variation bands as  defined in the text. 
}}
\end{figure}

In Fig.\,\ref{fig3} we consider the ratio of $q_T$ distributions
for $Z/\gamma^*$ and $W^\pm$ production and decay. We consider the quantity
\begin{equation}
    R(q_T) = \frac{\sigma_{Z}}{\sigma_{W}}\,\frac{d \sigma_{W}}{d q_T}\Big/\frac{d \sigma_{Z}}{d q_T},
\label{eq:RqT}
\end{equation}
where $\frac{1}{\sigma_{V}}\,\frac{d \sigma_{V}}{d q_T}$ with $V=W,Z$
is the normalized $q_T$ distribution for $W$ and $Z/\gamma^*$  production
and decay 
inclusive over the
leptonic final state kinematics, apart for a selection cut
on the invariant mass of the lepton pair: $80<M_{l^+ l^-}<100\,$GeV
and $M_{l\nu}>50\,$GeV.

In Fig.\,\ref{fig3} we show the
resummed component  of the transverse-momentum
distribution of Eq.\,\ref{eq:RqT} for the ratio $W^+/Z$ (left panel) and $W^-/Z$ (right panel) in the small-$q_T$ region.
From the results of Fig.\,\ref{fig3} (left and right panels) we observe that the scale dependence is greatly reduced (roughly by one order of magnitude) with respect to the distributions shown in  Figs.\,\ref{fig1},\ref{fig2}. 
The scale variation at  N$^4$LL+N$^4$LOa accuracy is around $\pm 0.3\%-0.4\%$  at  $q_T\sim 1\,$GeV, then it reduces at $\pm 0.1\%$  level at the peak ($q_T\sim 4\,$GeV), 
it further decrease  below $0.05\%$ level for $q_T\sim 7\,$GeV and then it slightly increase up to $\pm 0.2\%$ for $q_T\sim 30\,$GeV. 
This reduction of scale uncertainty is not unexpected because 
in the ratio correlated uncertainties on $W$ and $Z$ distributions cancel.
In particular higher order QCD predictions for the resummed component
of the cross section has a high degree of universalities
and the process dependence is mainly due to the different flavour content
of the partonic subprocesses for $W$ and $Z$ production.

One may wonder if correlated scale variation for the ratio of $W$ and $Z$
distribution can underestimate the true perturbative uncertainty. 
However the overlap of the scale uncertainty band indicates that
correlated scale variation at NLL+NLO, NNLL+NNLO and N$^3$LL+N$^3$LO
correctly estimate the size of higher-order corrections.
An alternative, and more robust, perturbative uncertainty can be obtained considering the size of the difference between the prediction at 
a given order with respect to the prediction at the previous order. 
In this way we obtain an uncertainty which is even smaller than the one obtained
through the perturbative scale variation method.

However we stress that the predictions presented in Fig.\,\ref{fig3} are far
from being complete since at such level of
theoretical precision several effects cannot be neglected. In particular
also very small effects which however are different in the $W$ and $Z$ case can give not negligible
effects on the $W/Z$ ratio. For instance the impact of the {\itshape process dependent} finite component
of the cross section, the (flavour dependent) non-perturbative 
{\itshape intrinsic $k_T$} effects\cite{Signori:2013mda}, the QED and electroweak effects\,\cite{Barze:2012tt,Barze:2013fru,Alioli:2016fum,Cieri:2018sfk,Autieri:2023xme},
the heavy-quark mass effects\,\cite{Bagnaschi:2018dnh,Pietrulewicz:2017gxc}.

In conclusion, in this paper we have presented the implementation of the $q_T$ resummation
formalism of Refs.\,\cite{Bozzi:2005wk,Bozzi:2010xn,Catani:2015vma}
for Drell--Yan processes up to N$^4$LL+N$^4$LO {\itshape approximated} accuracy in the
{\tt DYTurbo} numerical program\,\cite{Camarda:2019zyx,dyturbo}.
We have illustrated explicit numerical results
for the resummed component of the transverse-momentum
distribution for the case of $Z/\gamma^*, W^\pm$ 
production and leptonic decay at LHC energies.
We also considered theoretical predictions for the ratio of $W^\pm$ and $Z/\gamma^*$ $q_T$ distributions.
 Perturbative uncertainties have been
estimated through a study of the scale variation band.

The {\tt DYTurbo} numerical code allows the user to apply arbitrary kinematical cuts on the vector boson and the final-state leptons,
and to compute the corresponding relevant distributions in the form of bin histograms. These features make the {\tt DYTurbo} a useful tool for 
Drell--Yan studies at hadron colliders such as the Tevatron and the LHC. 

\section*{Acknowledgments} 
LC is supported by the Generalitat Valenciana (Spain) through the plan GenT program (CIDEGENT/2020/011) and his work is supported by the Spanish Government (Agencia Estatal de Investigación) and ERDF funds from European Commission (Grant no. PID2020-114473GB-I00 funded by MCIN/AEI/10.13039/501100011033).

\clearpage

\appendix
\section{Transverse-momentum resummation up to N$^4$LL+N$^4$LO  accuracy}
\label{appendix}

We consider the process
\beqn
\label{eq1}
h_1 + h_2 \to V +X \to l_3+l_4+X,
\eeqn
where $V$ denotes the vector boson
produced
by the colliding hadrons $h_1$ and $h_2$ with a centre--of--mass energy $s$,
while $l_3$ and $l_4$ are the final state leptons produced by the $V$ decay.
The lepton kinematics
is completely specified in terms of  the transverse-momentum ${\bf q_T}$
(with $q_T=\sqrt{{\bf q_T}^2}$), the rapidity $y$
and the invariant
mass $M$ of the lepton pair, and by two additional variables ${\bf \Omega}$
that specify the angular distribution of the leptons
with respect to the vector boson momentum.

We consider the Drell--Yan  cross section
fully differential in the leptonic final state. According to the factorization theorem we can write
\beqn
\label{diffXS}
    \frac{d {\sigma}_{h_1h_2\to l_3l_4}}{d^2{\bf q_T}dM^2dy d\bf{\Omega}}
    ({\bf q_T},M^2,y, s, {\bf\Omega})
&=&
\sum_{a_1,a_2}\int_0^1dx_1\int_0^1dx_2 
\,f_{a_1/h_1}( x_1,\mu_F^2)\,f_{a_2/h_2}(x_2,\mu_F^2) \nonumber \\
&\times&\,
{\frac{d{\hat\sigma}_{a_1a_2\to l_3l_4}}{d^2{\bf q_T}\,d{M^2}\,d\hat y\,d{\bf \Omega}}} 
({\bf q_T},M,\hat y,\hat s,{\bf \Omega};\alpha_S,\mu_R^2,\mu_F^2)\,,
    \eeqn
    where $f_{a/h}(x,\mu_F^2)$ ($a=q_f,\bar q_f,g$) are the parton distribution functions
    of the hadron $h$, 
$\hat s = x_1 x_2 s$ is
the partonic  centre--of--mass energy squared, 
$\hat y=y-\ln\sqrt{x_1/x_2}$ is the vector boson rapidity with respect to the
colliding partons while $\mu_R$ and  $\mu_F$
are the renormalization and factorization scales. 
The last factor in the right-hand side of Eq.\,(\ref{diffXS})  is
multi-differential partonic cross sections,
computable in perturbative QCD as a series expansion in the strong coupling $\alpha_S=\alpha_S(\mu_R)$,
which will be denoted in the following  by the shorthand notation $[d \hat\sigma_{a_1a_2\to l_3l_4}]$.

The partonic cross section can be decomposed as
\beqn
\label{partXS2}
    [d \hat\sigma_{a_1a_2\to l_3l_4}]= [d \hat\sigma^{({\rm res.})}_{a_1a_2\to l_3l_4}]+
    [d \hat\sigma^{({\rm fin.})}_{a_1a_2\to l_3l_4}]
\eeqn
where the
first term on the right-hand side of Eq.\,(\ref{partXS2})
is the resummed component
which dominates in the small $q_T$ region
while the second term is the finite component which is needed
at large $q_T$.

We briefly review the impact-parameter space $b$\,\cite{Parisi:1979se}
resummation formalism of Refs.\,\cite{Bozzi:2005wk,Bozzi:2010xn,Catani:2015vma}. The resummed 
component in the r.h.s. of Eq.\,\ref{partXS2} can then be written as 
\beqn
\label{resum}
\left[ {d{\hat \sigma}_{a_1a_2\to l_3l_4}^{(\rm res.)}} \right]
= \sum_{b_1,b_2=q,{\bar q}} \!\!
\frac{{d{\hat \sigma}^{(0)}_{b_1b_2\to l_3l_4}}}{d{\bf{\Omega}}} \;
\frac{1}{\hat s} \;
\int_0^\infty \frac{db}{2\pi} \; b \,J_0(b q_T) 
\;{\cal W}_{a_1a_2,b_1b_2\to V}(b,M,\hat y,\hat s;\alpha_S,\mu_R^2,\mu_F^2) \;,
\eeqn
where
$J_0(x)$ is the $0$th-order Bessel function and 
the factor $d{\hat \sigma}^{(0)}_{b_1b_2\to l_3l_4}$ 
is the Born level differential cross section for the partonic subprocess
$q\bar q\to V \to l_3l_4$.

The function
${\cal W}_{V}(b,M,\hat y,\hat s)$ 
can be expressed in an 
exponential form
by considering 
the `double' $(N_1,N_2)$ Mellin moments 
with respect to the variables 
$z_1=e^{+\hat y}M/{\sqrt{\hat s}}$ and $z_2=e^{-\hat y}M/{\sqrt{\hat s}}$ 
at fixed $M$\,\footnote{
For the sake of simplicity in our symbolic notation
the explicit dependence  on parton indices (which are relevant for the exponentiation in the multiflavour space)
and the double Mellin indices are understood.
The interested reader can find the details in Ref.\,\cite{Bozzi:2005wk} (in particular  Appendix A) and Ref.\cite{Bozzi:2007pn}.}
\cite{Bozzi:2005wk,Bozzi:2007pn}
\begin{equation}
\label{wtilde}
{\cal W}_{V}(b,M;\alpha_S,\mu_R^2,\mu_F^2)
={\cal H}_V\left(
\alpha_S;M/\mu_R,M/\mu_F,M/Q \right) 
\times \exp\{{\cal G}(\alpha_S,L;M/\mu_R,M/Q)\}\,,
\end{equation}
where 
we have introduced
the logarithmic expansion parameter
\begin{equation}
  L\equiv \ln ({Q^2 b^2}/{b_0^2})
\end{equation}  
with $b_0=2e^{-\gamma_E}$ ($\gamma_E=0.5772...$ 
is the Euler number).
The scale $Q\sim M$ 
is the resummation scale \cite{Bozzi:2003jy}, 
which parameterizes the
arbitrariness in the resummation procedure.

The process dependent function ${\cal H}_V(\alpha_S)$\,\cite{Catani:2000vq,Catani:2013tia}
includes the hard-collinear contributions 
and it can be written in term of a process dependent hard factor $H_V(\alpha_S)$
and two process independent functions ${C}(\alpha_S)$ associated to collinear emissions
from the initial state 
colliding partons\,\footnote{A simple specification of a resummation scheme
customarily used in the literature on $q_T$ resummation for vector boson 
is: $H_V(\alpha_S) \equiv 1$ (i.e.\ $H_V^{(n)}=0$ for $n>0$).}
\begin{equation}
\label{hcc}
      {\cal H}_V(\alpha_S)=
      H_V(\alpha_S)\,{C}(\alpha_S)\,{C}(\alpha_S)\,.
\end{equation}
The functions in Eq.(\ref{hcc}) have  a standard perturbative expansion 
\begin{eqnarray}
\label{hskew}
{\cal H}_V(\alpha_S)&=&1+\sum_{n=1}^{\infty}\left( \frac{\alpha_S}{\pi}\right)^n\,{\cal H}_V^{(n)}, \\
\label{hhard}
H_V(\alpha_S)&=&1+\sum_{n=1}^{\infty}\left( \frac{\alpha_S}{\pi}\right)^n\,{H}_V^{(n)}, \\
\label{cfun}
C(\alpha_S)&=&1+\sum_{n=1}^{\infty}\left( \frac{\alpha_S}{\pi}\right)^n\,{C}^{(n)}\,,
\end{eqnarray}
therefore up to the fourth order we have the following relations
\begin{eqnarray}
{\cal H}_V^{(1)}&=& {H}_V^{(1)} +{C}^{(1)}+{C}^{(1)},\\
{\cal H}_V^{(2)}&=& {H}_V^{(2)} +{C}^{(2)}+{C}^{(2)}+{H}_V^{(1)}({C}^{(1)}+{C}^{(1)})+{C}^{(1)}{C}^{(1)},\\
{\cal H}_V^{(3)}&=& {H}_V^{(3)} +{C}^{(3)}+{C}^{(3)}+{H}_V^{(2)}({C}^{(1)}+{C}^{(1)})+{H}_V^{(1)}({C}^{(2)}+{C}^{(2)}+{C}^{(1)}{C}^{(1)})
\nonumber\\
&+&{C}^{(2)}{C}^{(1)}+{C}^{(2)}{C}^{(1)},\\
{\cal H}_V^{(4)}&=& {H}_V^{(4)} +{C}^{(4)}+{C}^{(4)}+{H}_V^{(3)}({C}^{(1)}+{C}^{(1)})+{H}_V^{(2)}({C}^{(2)}+{C}^{(2)}+{C}^{(1)}{C}^{(1)})\nonumber\\
&+&{H}_V^{(1)}({C}^{(3)}+{C}^{(3)}+{C}^{(2)}{C}^{(1)}+{C}^{(2)}{C}^{(1)})+{C}^{(3)}{C}^{(1)}+{C}^{(3)}{C}^{(1)}+{C}^{(2)}{C}^{(2)}\,.
\end{eqnarray}

The universal (process independent) form factor $\exp\{{\cal G}\}$ 
in the right-hand side of Eq.~(\ref{wtilde})
contains all
the terms that order-by-order in $\alpha_S$ are logarithmically divergent 
as $b \to \infty$ (i.e.\ $q_T\to 0$). 
The resummed logarithmic expansion of ${\cal G}$ reads\,\cite{Bozzi:2005wk}
\begin{eqnarray}
\label{exponent} 
      {\cal G}(\alpha_S,L)&=&-\int_{b_0^2/b^2}^{Q^2}\frac{dq^2}{q^2}\left[A(\alpha_S(q^2))\ln\frac{M^2}{q^2}+\widetilde B(\alpha_S(q^2))\right]\nonumber\\
      &=&      L\, g^{(1)}(\alpha_S L)+g^{(2)}(\alpha_S L)+\sum_{n=1}^{\infty}\left( \frac{\alpha_S}{\pi}\right)^n\,{g}^{(n+2)}(\alpha_S L)\,,
\end{eqnarray}
where the functions $g^{(n)}$ control and resum the $\alpha_S^kL^{k}$ (with $k\geq 1$) logarithmic terms in the exponent of Eq.\,(\ref{wtilde})
due to soft and collinear radiation.
The perturbative functions $A(\alpha_S)$ and $\widetilde B(\alpha_S)$  can be expanded as
\begin{eqnarray}
A(\alpha_S)&=&\sum_{n=1}^{\infty}\left( \frac{\alpha_S}{\pi}\right)^n\,{A}^{(n)}\,,\\
\widetilde B(\alpha_S)&=&\sum_{n=1}^{\infty}\left( \frac{\alpha_S}{\pi}\right)^n\,{\widetilde B}^{(n)}\,.\\
\end{eqnarray}
The function $\widetilde B(\alpha_S)$ can be written as follows
\begin{eqnarray}
  \label{Btilde}
\widetilde B(\alpha_S)&=&B(\alpha_S) +2\beta(\alpha_S) \frac{d\ln C(\alpha_S)}{d\ln\alpha_S}+2\gamma(\alpha_S)\,,
\end{eqnarray}
in terms of the resummation coefficient $B(\alpha_S)$, the collinear functions  $C(\alpha_S)$ (see Eq.(\ref{cfun})), the functions $\gamma(\alpha_S)$
(the Mellin moments of
the Altarelli--Parisi splitting functions
\,\footnote{
In order to match the effect of the charm and bottom-mass threshold included in the evolution of PDFs in Eq.(\ref{diffXS}), the resummation (evolution) effects due to the $\gamma(\alpha_S)$ term in Eq.(\ref{exponent}) are asymptotically switched off 
when approaching their corresponding quark-mass thresholds through a $b_\star$ prescription (see Eq.(\ref{bstar})) with values of $b_{\textrm{lim}}=m_q$.}
) and the QCD $\beta$ function
\begin{eqnarray}
\label{beta}
\frac{d\ln \alpha_S(\mu^2)}{d\ln\mu^2}=\beta(\alpha_S)=-\sum_{n=0}^{+\infty}\beta_n\left(\frac{\alpha_S}{\pi}\right)^{n+1}\,.
\end{eqnarray}

By explicit integration of Eq.(\ref{exponent}) we obtain the following $g^{(i)}$ for $1\leq i \leq 5$
\begin{align}
\label{g1fun}
g^{(1)}(\as L) &= \f{A^{(1)}}{\beta_0} \f{\la+\ln(1-\la)}{\la} \;\;,  \\
~\nonumber\\
\label{g2fun}
g^{(2)}(\as L) 
&= \f{{\overline B}^{(1)}}{\beta_0}
\ln(1-\la) -\f{A^{(2)}}{\beta_0^2} 
\left( \f{\la}{1-\la} +\ln(1-\la)\right) \nn \\ 
&+ \f{A^{(1)}}{\beta_0} 
\left( \f{\la}{1-\la} +\ln(1-\la)\right) \ln\f{Q^2}{\mu_R^2}  \nn \\
& +\f{A^{(1)} \beta_1}{\beta_0^3} \left( \f{1}{2} \ln^2(1-\la)+ 
\f{\ln(1-\la)}{1-\la} + \f{\la}{1-\la}  \right) \;,
\end{align}

\begin{align}
\label{g3fun}
g^{(3)} (\as L) 
&= -\f{A^{(3)}}{2 \beta_0^2} \f{\la^2}{(1-\la)^2}
-\f{{\overline B}^{(2)}}{\beta_0} \f{\la}{1-\la}  
+\f{A^{(2)} \beta_1}{\beta_0^3} \left( \f{\la (3\la-2)}{2(1-\la)^2}
 - \f{(1-2\la) \ln(1-\la)}{(1-\la)^2} \right) \nn \\
& + \f{{\overline B}^{(1)} \beta_1}{\beta_0^2} \left( \f{\la}{1-\la} + 
\f{\ln(1-\la)}{1-\la} \right) - \f{A^{(1)}}{2} \f{\la^2}{(1-\la)^2}  
\ln^2\f{Q^2}{\mu_R^2} \nn \\
&+ \ln\f{Q^2}{\mu_R^2} \left( {\overline B}^{(1)} \f{\la}{1-\la} + 
\f{A^{(2)}}{\beta_0}
 \f{\la^2}{(1-\la)^2} + A^{(1)} \f{\beta_1}{\beta_0^2}
 \left( \f{\la}{1-\la} + \f{1-2\la}{(1-\la)^2} \ln(1-\la) \right) \right) 
 \nn \\
&   +A^{(1)} \left( \f{\beta_1^2}{2 \beta_0^4} \f{1-2\la}{(1-\la)^2} 
\ln^2(1-\la) 
+ \ln(1-\la) \left[  \f{\beta_0 \beta_2 -\beta_1^2}{\beta_0^4} 
+\f{\beta_1^2}{\beta_0^4 (1-\la)}  \right] \right.  \nn \\
& \left. + \f{\la}{2 \beta_0^4 (1-\la)^2} ( \beta_0 \beta_2 (2-3\la)+
\beta_1^2 \la) \right) \;,       
\end{align}
\begin{align}
\label{g4fun}
g^{(4)}(\as L) &= -\f{A^{(4)}}{6\beta_0^2}\f{(3-\la)\la^2}{(1-\la)^3}-\f{{\overline B}^{(3)}}{2\beta_0}\frac{(2-\la)\la}{(1-\la)^2}-\f{A^{(3)}}{2\beta_0}\Bigg(\f{\beta_1}{\beta_0^2}\bigg[
\f{(6-15\la+5\la^2)\la}{6(1-\la)^3}\;\;\nn  \\
&+\f{(1-3\la)}{(1-\la)^3}\ln(1-\la) \bigg] - \f{(3-\la)\la^2}{(1-\la)^3}\ln\f{Q^2}{\mu_R^2} \Bigg)+{\overline B}^{(2)}\Bigg(\f{\beta_1}{\beta_0^2}\bigg[\f{(2-\la)\la}{2(1-\la)^2}+\f{\ln(1-\la)}{(1-\la)^2} \bigg] \;\;\nn  \\
&+\f{(2-\la)\la}{(1-\la)^2}\ln\f{Q^2}{\mu_R^2}\Bigg)+A^{(2)}\Bigg( -\f{2\beta_2}{3\beta_0^3}\f{\la^3}{(1-\la)^3}+\f{\beta_1^2}{2\beta_0^4}\bigg(\f{\la(6-9\la+11\la^2)}{6(1-\la)^3}+\f{\ln(1-\la)}{(1-\la)^2} \;\;\nn  \\
&+\f{1-3\la}{(1-\la)^3}\ln^2(1-\la)\bigg)+\bigg[\f{\beta_1}{2\beta_0^2}\f{(2-\la)\la}{(1-\la)^2}+\f{\beta_1}{\beta_0^2}\f{1-3\la}{(1-\la)^3}\ln(1-\la) \bigg]\ln\f{Q^2}{\mu_R^2}\;\;\nn  \\
&-\f{(3-\la)\la^2}{2(1-\la)^3}\ln^2\f{Q^2}{\mu_R^2}\Bigg)+{\overline B}^{(1)}\Bigg(\f{\beta_1^2}{2\beta_0^3}\bigg[\f{\la^2}{(1-\la)^2}-\f{\ln^2(1-\la)}{(1-\la)^2} \bigg]-\f{\beta_2}{2\beta_0^2}\f{\la^2}{(1-\la)^2}\;\;\nn  \\
&-\f{\beta_1}{\beta_0}\f{\ln(1-\la)}{(1-\la)^2}\ln\f{Q^2}{\mu_R^2}-\f{\beta_0}{2}\f{(2-\la)\la}{(1-\la)^2}\,\ln^2\f{Q^2}{\mu_R^2}\Bigg)+A^{(1)}\Bigg(-\f{\beta_1^3}{\beta_0^5}\bigg(\f{\la^3}{6(1-\la)^3}\;\;\nn  \\
&+\f{(1+\la)\la^2}{2(1-\la)^3}\ln(1-\la)+\f{\la}{2(1-\la)^3}\ln^2(1-\la) + \f{(1-3\la)}{6(1-\la)^3}\ln^3(1-\la) \bigg)\;\;\nn  \\
&-\f{\beta_1\beta_2}{2\beta_0^4}\bigg( \f{\la(6-15\la+5\la^2)}{6(1-\la)^3}+\f{(1-3\la+2\la^2-2\la^3)}{(1-\la)^3} \ln(1-\la)\bigg)\;\;\nn  \\
&+\f{\beta_3}{2\beta_0^3}\bigg(\f{\la(6-15\la+7\la^2)}{6(1-\la)^3}+\ln(1-\la) \bigg)+\Bigg[-\f{\beta_1^2}{\beta_0^3}\bigg(\f{\la^2(1+\la)}{2(1-\la)^3}+\f{\la}{(1-\la)^3}\ln(1-\la)\;\;\nn  \\
&+\f{1-3\la}{2(1-\la)^3}\ln^2(1-\la)\bigg)+\f{\beta_2}{2\beta_0^2}\f{\la^2(1+\la)}{(1-\la)^3}\Bigg]\ln\f{Q^2}{\mu_R^2}-\f{\beta_1}{2\beta_0}\Bigg[\f{\la}{(1-\la)^3}\;\;\nn  \\
&+\f{1-3\la}{(1-\la)^3}\ln(1-\la) \Bigg]\ln^2\f{Q^2}{\mu_R^2}+\f{\beta_0}{6}\f{(3-\la)\la^2}{(1-\la)^3}\ln^3\f{Q^2}{\mu_R^2}
\Bigg)\;,
\end{align}

\begin{align}
\label{g5fun}
g^{(5)}(\as L) &= -\f{A^{(5)}}{12\beta_0^2} \f{\la^2 (6-4 \la + \la^2)}{(1-\la)^4} -  \f{{\overline B}^{(4)}}{3\beta_0} \f{\la (3-3 \la+\la^2)}{(1-\la)^3}\;\;\nn  \\
&+\f{A^{(4)}}{3\beta_0}\Bigg(\f{\beta_1}{\beta_0^2}\Bigg[\f{\la (-12+42\la-28\la^2+7\la^3)}{12(1-\la)^4} -\f{1-4\la }{(1-\la)^4}\ln(1-\la)\Bigg]
\;\;\nn\\ 
 &+\f{\la^2(6-4\la+\la^2)}{(1-\la)^4}\ln\f{Q^2}{\mu_R^2}\Bigg)+{\overline B}^{(3)}\Bigg( \f{\beta_1}{\beta_0^2}\Bigg[\f{\la (3-3\la+\la^2)}{3(1-\la)^3}+\f{\ln(1-\la)}{(1-\la)^3}\Bigg]\;\;\nn\\ 
 &+ \f{\la(3-3\la+\la^2)}{(1-\la)^3}\ln\f{Q^2}{\mu_R^2}  \Bigg) + A^{(3)}\Bigg(-\f{\beta_2}{4\beta_0^3} \f{\la^3 (4-\la)}{(1-\la)^4}\;\;\nn\\
 &+ \f{\beta_1^2}{\beta_0^4} \Bigg[\f{\la (12-24\la+52\la^2-13\la^3)}{36(1-\la)^4}+\f{\ln(1-\la)}{3(1-\la)^3}+\f{1-4\la}{2(1-\la)^4}\ln^2(1-\la) \Bigg]\;\;\nn\\
 &+\f{\beta_1}{\beta_0^2} \Bigg[\f{\la(3-3\la+\la^2)}{3(1-\la)^3} + \f{1-4\la}{(1-\la)^4}\ln(1-\la) \Bigg] \ln\f{Q^2}{\mu_R^2} \;\;\nn\\
 &-\f{\la^2(6-4\la+\la^2)}{2(1-\la)^4}\ln^2\f{Q^2}{\mu_R^2} \Bigg) + {\overline B}^{(2)}\Bigg( -\f{\beta_2}{3\beta_0^2} \f{(3-\la)\la^2}{(1-\la)^3} + \f{\beta_1^2}{\beta_0^3} \bigg(\f{(3-\la)\la^2}{3(1-\la)^3}\;\;\nn\\
 &- \f{\ln^2(1-\la)}{(1-\la)^3}\bigg) - \f{2\beta_1}{\beta_0} \f{\ln(1-\la)}{(1-\la)^3}\ln\f{Q^2}{\mu_R^2} -\beta_0 \f{\la(3-3\la+\la^2)}{(1-\la)^3}\ln^2\f{Q^2}{\mu_R^2}  \Bigg)\;\;\nn\\
 &+ A^{(2)}\Bigg(-\f{\beta_3}{12\beta_0^3} \f{\la^3(8-5\la)}{(1-\la)^4} + \f{\beta_1\beta_2}{3\beta_0^4} \bigg(\f{\la (6-21\la+44\la^2-20\la^3)}{6(1-\la)^4} \;\;\nn\\
 &+\f{1-4\la+9\la^2}{(1-\la)^4} \ln(1-\la)\bigg)+\f{\beta_1^3}{\beta_0^5}\bigg(\f{\la(-12+42\la-64\la^2+25\la^3)}{36(1-\la)^4} \;\;\nn\\
 &-\f{(1-4\la+9\la^2)}{3(1-\la)^4} \ln(1-\la)-\f{\la}{(1-\la)^4} \ln^2(1-\la)- \f{1-4\la}{3(1-\la)^4}\ln^3(1-\la)\bigg)\;\;\nn\\
 &+\Bigg[\f{\beta_2}{3\beta_0^2} \f{(3+4\la-\la^2)\la^2}{(1-\la)^4}+\f{\beta_1^2}{\beta_0^3}\bigg( -\f{(3+4\la-\la^2)\la^2}{3(1-\la)^4} - \f{2\la}{(1-\la)^4} \ln(1-\la)\;\;\nn\\
 &-\f{1-4\la}{(1-\la)^4}\ln^2(1-\la)\bigg) \Bigg]\ln\f{Q^2}{\mu_R^2} + \f{\beta_1}{\beta_0} \Bigg[ -\f{\la}{(1-\la)^4} - \f{1-4\la}{(1-\la)^4} \ln(1-\la) \Bigg]\ln^2\f{Q^2}{\mu_R^2}\;\;\nn\\
 &+ \f{\beta_0}{3} \f{\la^2(6-4\la+\la^2)}{(1-\la)^4}\ln^3\f{Q^2}{\mu_R^2} \Bigg)+{\overline B}^{(1)}\Bigg(-\f{\beta_3}{6\beta_0^2} \f{(3-2\la)\la^2}{(1-\la)^3} + \f{\beta_1 \beta_2}{\beta_0^3} \bigg(\f{(3-2\la)\la^2}{3(1-\la)^3} \;\;\nn\\
 &+ \f{\la}{(1-\la)^3} \ln(1-\la)\bigg)+\f{\beta_1^3}{\beta_0^4} \bigg( -\f{(3-2\la)\la^2}{6(1-\la)^3} - \f{\la}{(1-\la)^3} \ln(1-\la) - \f{\ln^2(1-\la)}{2(1-\la)^3} \;\;\nn\\
 &+ \f{\ln^3(1-\la)}{3(1-\la)^3} \bigg)+ \Bigg[\f{\beta_2}{\beta_0} \f{\la}{(1-\la)^3} + \f{\beta_1^2}{\beta_0^2} \bigg(-\f{\la}{(1-\la)^3}- \f{\ln(1-\la)}{(1-\la)^3}+ \f{\ln^2(1-\la)}{(1-\la)^3}\bigg) \Bigg]\ln\f{Q^2}{\mu_R^2}\;\;\nn\\
 &+\beta_1 \Bigg[-\f{\la(3-3\la+\la^2)}{2(1-\la)^3}+\f{\ln(1-\la)}{(1-\la)^3} \Bigg]\ln^2\f{Q^2}{\mu_R^2} + \beta_0^2 \f{\la(3-3\la+\la^2)}{3(1-\la)^3} \ln^3\f{Q^2}{\mu_R^2}  \Bigg)\;\;\nn\\
 &+A^{(1)}\Bigg(\f{\beta_2^2}{3\beta_0^4}\bigg( \f{\la(-12+42\la-52\la^2+7\la^3)}{12(1-\la)^4}-\ln(1-\la) \bigg) \;\;\nn
\end{align}

\begin{align}
     &+\f{\beta_4}{3\beta_0^3}\bigg( \f{\la(12-42\la+40\la^2-13\la^3)}{12(1-\la)^4}+\ln(1-\la) \bigg)+\f{\beta_1\beta_3}{6\beta_0^4}\bigg(-\f {\la (2 - 5 \la)} {3}\f { (3  -3\la + \la^2)} {(1 - \la)^4}\;\;\nn\\
 & -\f{2-8\la+9\la^2-10\la^3+4\la^4}{(1-\la)^4} \ln(1-\la)\bigg) + \f{\beta_1^2\beta_2}{\beta_0^5}\bigg(\f{\la(12-42\la+52\la^2+5\la^3)}{36(1-\la)^4} \;\;\nn\\
 &-\f{(-1+3\la-3\la^2+3\la^3)}{3(1-\la)^3}\ln(1-\la)-\f{3\la^2}{2(1-\la)^4}\ln^2(1-\la)\bigg) + \f{\beta_1^4}{2\beta_0^6} \bigg(-\f{\la^3(2+3\la)}{6(1-\la)^4} \;\;\nn\\
 &+\f {\la^2 (-3 + 2\la - 2\la^2)} {3 (1 - \la)^4}\ln(1-\la)-\f{(1-3\la)\la}{(1-\la)^4}\ln^2(1-\la)-\f{1-6\la}{3(1-\la)^4}\ln^3(1-\la)\;\;\nn\\
 &+\f{1-4\la}{6(1-\la)^4}\ln^4(1-\la) \bigg)+ \Bigg[-\f{\beta_3}{6\beta_0^2}\f{\la^2(-3-2\la+2\la^2)}{(1-\la)^4} - \f{\beta_1 \beta_2}{\beta_0^3}\bigg( \f{2\la^3}{3(1-\la)^3} + \f{3\la^2}{(1-\la)^4} \ln(1-\la)\bigg) \;\;\nn\\
 &+\f{\beta_1^3}{\beta_0^4}\bigg(-\f{\la^2(3-2\la+2\la^2)}{6(1-\la)^4}-\f{(1-3\la)\la}{(1-\la)^4}\ln(1-\la) - \f{1-6\la}{2(1-\la)^4} \ln^2(1-\la)\;\;\nn\\
 &+\f{1-4\la}{3(1-\la)^4} \ln^3(1-\la) \bigg)
 \Bigg]\ln\f{Q^2}{\mu_R^2} + \Bigg[-\f{3\beta_2}{2\beta_0}\f{\la^2}{(1-\la)^4} + \f{\beta_1^2}{2\beta_0^2} \bigg( -\f{(1-3\la)\la}{(1-\la)^4}- \f{(1-6\la)}{(1-\la)^4} \ln(1-\la)\;\;\nn\\ 
 &+\f{(1-4\la)}{(1-\la)^4}\ln^2(1-\la) \bigg)\Bigg]\ln^2\f{Q^2}{\mu_R^2} + \f{\beta_1}{3}\Bigg[ \f{\la(2+6\la-4\la^2+\la^3)}{2(1-\la)^4}+ \f{1-4\la}{(1-\la)^4} \ln(1-\la)\Bigg]\ln^3\f{Q^2}{\mu_R^2} \;\;\nn\\
 &- \f{\beta_0^2}{12}\f{(6-4\la+\la^2)\la^2}{(1-\la)^4}\ln^4\f{Q^2}{\mu_R^2}\Bigg)\;,
\end{align}
where
\begin{equation}
\lambda = \frac{1}{\pi} \,\beta_0 \,\as(\mu_R^2) \,L \;\;,
\end{equation}
\begin{equation}
{\overline B}^{(n)} = {\widetilde B}^{(n)} + 
A^{(n)} \ln\f{M^2}{Q^2}\;\;.
\end{equation}
The $g^{(1)}$, $g^{(2)}$ and $g^{(3)}$ resummation functions can be found in Ref.\,\cite{Bozzi:2005wk}. The $g^{(4)}$ function can be found in Ref.\,\cite{Bizon:2017rah} for the related case of direct transverse momentum space resummation. The explicit expression of the first five coefficients of the $\beta$ function, can be found in the following references: $\beta_0$, $\beta_1$ and $\beta_2$ in Refs. \cite{Tarasov:1980au,Larin:1993tp}, $\beta_3$ in Ref. \cite{vanRitbergen:1997va} and $\beta_4$ in \cite{Herzog:2017ohr}.

At NLL+NLO we include the functions $g^{(1)}$, $g^{(2)}$ and ${\cal H}_V^{(1)}$, 
at NNLL+NNLO we also include the functions $g^{(3)}$
and ${\cal H}_V^{(2)}$\,\cite{Catani:2012qa,Gehrmann:2012ze}, at 
N$^3$LL+N$^3$LO  the functions $g^{(4)}$ and ${\cal H}_V^{(3)}$\,\,\cite{Luo:2019szz,Ebert:2020yqt} and finally at
N$^4$LL+N$^3$LO  the function $g^{(5)}$ and ${\cal H}_V^{(4)}$.

\begin{figure}[t]
\begin{center}
  \includegraphics[width=0.7\textwidth]{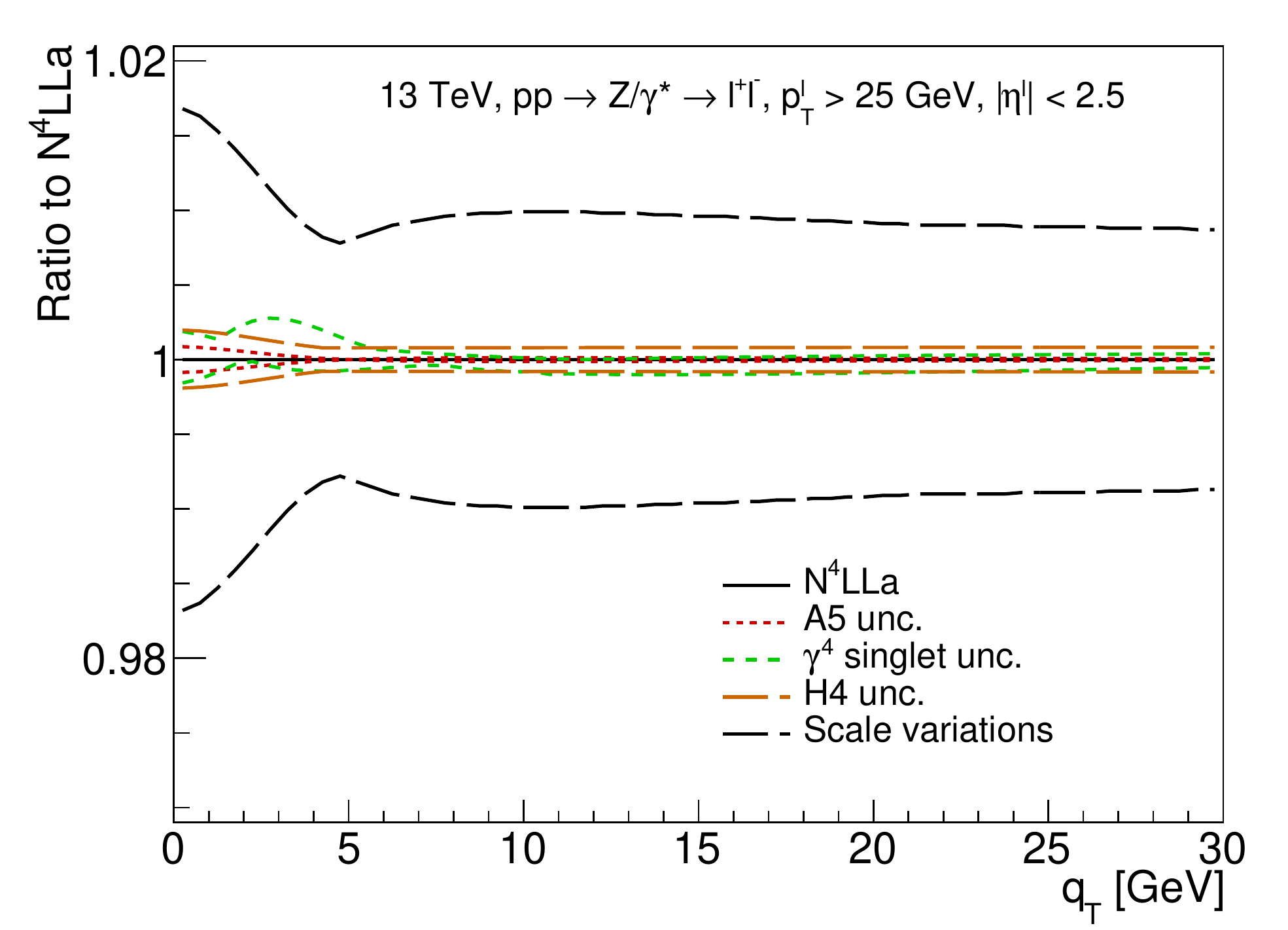}
\end{center}
\caption
{ \label{fig:unc}
  { \em Uncertainties arising from numerical approximations or incomplete knowledge of the perturbative coefficients at N$^4$LL+N$^4$LOa, compared to missing higher order uncertainties estimated with scale variations at this order.}}
\end{figure}

We consider uncertainties in the numerical approximations of the N$^4$LL coefficients, and estimate uncertainties arising from the incomplete knowledge of the N$^4$LO perturbative coefficients. The $B^{(4)}$ coefficient and the non-singlet four-loop splitting functions are known with good numerical approximation~\cite{Das:2019btv,Moult:2022xzt,Moch:2017uml}, the corresponding relative uncertainties on the $q_T$ distribution are at the level of $10^{-6}$ or smaller, and considered negligible. The numerical approximations of $A^{(5)}$~\cite{Herzog:2018kwj,Henn:2019swt,vonManteuffel:2020vjv,Li:2016ctv,Vladimirov:2016dll,Moch:2004pa,Li:2014afw} and of the 4-loop singlet splitting functions~\cite{Moch:2021qrk,Falcioni:2023luc} are the dominant uncertainties in the N$^4$LL approximation, and they amount to $1$--$3 \cdot 10^{-3}$ relative uncertainty. In order to estimate the size of the  
unknown the  $C^{(4)}$ coefficients \cite{Lee:2022nhh} we perform a Levin transform of the corresponding perturbative series~\cite{David:2013gaa,Levin1972DevelopmentON} to guess the value of the fourth term in these series, and assign to it a $100\%$ uncertainty. This is equivalent to assuming that the Levin transform is able to estimate the sign and the order of magnitude of these unknown coefficients. The corresponding uncertainty is at the level of $1$--$2 \cdot 10^{-3}$, and affects mostly the overall normalization. The uncertainties in the N$^4$LL+N$^4$LO approximation are shown in Fig.~\ref{fig:unc}, and found to be 5 to 10 times smaller compared to the missing higher order uncertainties estimated through scale variations.

\clearpage
\bibliography{dyqtN4LL}

\end{document}